# Straub tail, the deprivation effect and addiction to aggression

Natalia N. Kudryavtseva

Neurogenetics of Social Behavior Sector, Institute of Cytology and Genetcis SD RAS,

Novosibirsk, Russia, natnik@bionet.nsc.ru

Abstract

**Background.** It has been observed that male mice who are consistently winning fights with conspecifics in the settings of the sensory contact model can for no apparent reason raise their tail, which is very similar to a morphine-induced Straub tail response. Since this response is a typical index of opiate activation, it has been proposed that the opioidergic systems of such mice are chronically activated. This activation appeared to be a potent factor, which leads to addiction to aggression; in which case the subject is motivated to repeatedly display. To check this hypothesis, we exposed the mice who had won 20 fights in succession with conspecifics to a behavioral sensitization procedure.

**Material/Methods.** Male mice who engage in aggression against conspecifics and win were identified using the sensory contact model. The effects of the mu-opioid receptor agonist morphine (10 mg/kg, i.p.) on their behavior were examined in an open-field test before and after 5- and 14-day deprivation of aggression. Additionally, aggression levels (latency to the first attack, the number of attacks and the total attack duration) were measured before and after deprivation during a 10-min agonistic interaction and then compared.

**Results.** Morphine had a much stronger stimulating effect on the open-field behavior of 60 % of the winners deprived of aggression for 14 days than on that of the control mice. Morphine did not stimulate behavioral activity in the winners before or after deprivation for 5 days. Latency to raise the tail after morphine injection was significantly shorter in the winners than in the controls. The aggression level in the winners was higher after than before deprivation.

**Conclusion.** It has been concluded that, in the winners, the mu-opioid receptors became tolerant to the effects of the endogenous mu-opioids due to the activation of opioidergic systems in the brain and became sensitized after long aggression deprivation. The development of addiction to aggression due to repeated victories is discussed in the light of the theory of addiction proposed by Robinson and Berridge (2003).

**Introduction**

According to many [for review, see, for example, 1-5], aggression is rewarding and, like other basic behaviors, aggressive behavior in animals and humans is strongly influenced by the previous experience of aggression and any positive reinforcer can create a tendency to behave aggressively [3, 6, 7]. Rats and mice who have previously won fights in agonistic encounters attack more frequently in subsequent encounters [8-12]. Mice who have been repeatedly given the opportunity to display aggression engage in fights more frequently than those lacking such experience [13-15]. The same refers to humans: the individuals who once displayed aggressive behavior tend to do so again [3]. Psychologists note that, individuals who have a habit of behaving aggressively are most likely to follow the same pattern in behavior, when in a frustrating environment [for review, 6, 16].

Behavioral observations of mice exposed to the sensory contact model [17, 18] suggest that positive fighting experince gives as a permanent reward to the winners, hence a tendency to repeat aggression acts. Mice who have won 10 – 20 fights in succession display hostile behavior (attacks, threats, aggressive grooming, vigorous aggression) even towards a much heavier and stronger male [18]. They can keep attacking a defeated conspecific even though it displays a total submissiveness.

It has been observed that the winners can exhibit an elevated tail resembling a morphine-induced Straub tail [19], a typical symptom of opiate activation [20], in no matter what experimental conditions. Control animals did not display elevated tails in our experiments. Elevated tails and permanent rewarding from experiencing social victories suggest that the opioidergic systems of the winners are chronically activated, which is supported by our pharmacological data. It has been shown that the winners develop tolerance to the effects of the opioid receptor agonists or antagonists [21-24]. It has been proposed that, in the winners, there are dynamic changes in mu- and kappa-opioid receptors alongside the permanently activated opioid reward systems: these receptors may become desensitized (or sensitized, depending on the amount of positive fighting experience) to opioid drugs. It has also been hypothesized that repeated positive fighting experience can lead to addiction to engagement in aggression [25]. Experimental confirmation was obtained by exposure of 20-day winners to a sensitization procedure. The mu-opioid receptor agonist morphine was used for psychomotor stimulation, because evidence exists that endogenous morphine can function as a neuromodulator or a neurotransmitter in the CNS [26-28]. Also, morphine modulates aggressive behavior; its withdrawal enhances aggression in animals [29-33]. Additionally, the levels of aggression were measured in the winners before and after deprivation of aggression.

## Materials and methods

### Animals

Adult male C57BL/6J mice from our own stock were housed under standard conditions: a 12-h light/dark photocycle, food (pellets) and water available *ad libitum*. Mice were weaned at one month of age and housed in groups of 8-10 in plastic cages (36 x 23 x 12 cm). Experiments were performed on animals 10-12 weeks of age. All procedures were in compliance with the European Communities Council Directive No. 86/609/EEC of 24 November 1986.

### Sensory contact model

Male mice with repeated experiences of aggression leading to victories were generated using the sensory contact model [17]. Pairs of animals were placed in steel cages bisected by a perforated transparent partition, which allowed the animals to see, hear and smell their neighbor, but prevented physical contact. The animals were left undisturbed for two days for adaptation to new housing conditions (sensory contact) and then exposed to testing. Every day, at 14:00 – 17:00 local time, the steel lid was replaced by a transparent one and 5 min later, after adaptation to the new lighting conditions, the partition was removed for 10 min to facilitate agonistic interactions. The superiority of one of the mice was established within 2-3 test sessions. A superior mouse was attacking, biting and chasing another, who displayed only defensive behavior (sideways postures, upright postures, withdrawal, lying on the back or freezing). If intensive attacks lasted three minutes, the encounter was discontinued by lowering the partition. Each defeated mouse was then placed in a two-compartment cage with an unfamiliar winner behind the partition; the winners remained in their compartments. This procedure identified equal numbers of mice with opposite outcomes. The victorious mice were called "winners" and the defeated mice, "losers". Mice provided with individual housing for five days were used as controls (such mice appeared to be the most adequate controls for the sensory contact model: they lack submissiveness, normally displayed by most animals living as a group, while the effects of social isolation are still not there [17, 18].

Four animal groups were established: Controls, mice provided with individual housing for 5 days; Mice who won 20 fights and were not exposed to aggression deprivation (WIN); Mice who won 20 fights and were exposed to a short, 5-day period of aggression deprivation (WIN-SD); Mice who won 20 fights and were exposed to a long, 14-day period of aggression deprivation (WIN-LD).

During the deprivation periods, each winner was in the same cage as a 20-day loser behind the partition, which was kept lowered at all times.

**Drug treatment**

Animals in each group were treated either with saline or 20 mg/kg of the mu-opioid receptor agonist morphine hydrochloride (MOR) (0.1 ml/10 g body weight, i.p). Dosage was identified on the basis of literature data and verified for adequacy to the C57BL/6J mouse strain in preliminary experiments. Each treatment group consisted of 9 – 14 animals.

**Open-field test**

The open-field was an 80 x 80 cm plexiglas arena partitioned into 10 x 10 cm squares. A 150 W bulb was placed 150 cm above the field. The animals, each in its home cage, were brought to the experimental room. The steel lid of the cage was replaced by a transparent one for 5 min to facilitate arousal/activation. Each mouse was then placed individually in the center of the open field for 10 min for adaptation to the new lighting conditions. Each animal was then videotaped for 10 min, then given saline or MOR and videotaped during 30 min. The following behavioral variables were registered: 1) the number of crossed squares; 3) the number of rearings; 4) the duration of self-grooming (licking of the fur on the flanks or abdomen, washing over the head from ear to snout). The time interval between drug injection and tail elevation was measured three times for every animal and the mean of the two closest response times was used as the response time estimate. Between sessions, the open field was thoroughly washed with water and dried with napkins.

**Agonistic interactions**

The in-fight behavior of the 20-day winners was videotaped for 10 min on day 21. Some of these animals were then deprived of fighting opportunities for 5 days, the others, for 14 days. Latency to the first attack, the number of attacks, the total and average attack durations were measured before and after deprivation. These variables were then measured separately for the group of the winners who displayed increased total attack durations after deprivation.

**Statistical analysis**

The effects of membership in an experimental group (controls, WINs, WIN-SDs, WIN-LDs) on the number of rearings, crossed squares and the duration of self-grooming before drug and saline injections in the open-field test were estimated using one-way ANOVA. The effects of repeated positive fighting experience on behavioral variables were estimated by comparing data on the winners and the controls using the Student t-test for independent data. The attack variables in the winners before and after deprivation periods were estimated using the Student t-test for dependend data. After the rank transformation, the number of crossed squares over a 30-min period after injections was analysed using ANOVA with repeated measures (time intervals) and experimental group (controls, WINs, WIN-SDs, WIN-LDs) and treatment (saline and drug) as factors as well interaction between these factors. Whenever the interaction between these factors was determined, the Student t-test was applied. The effect of drug treatment was assessed by comparing drug and saline treatments using the Student t-test for independent data. A P value of ≤ 0.05 was considered statistically significant.

**Results**

**Basal behavioral activity in the open-field test before injections**

One-way ANOVA revealed a significant effect of experimental group membership on the total duration of self-grooming [$F(3,74) = 9.04$, $p < 0.001$] and on the number of crossed squares [$F(3,76) = 5.46$, $p < 0.01$], but not on the number of rearings [$F(3,75) = 2.24$, $p > 0.05$] before injections. An increased number of crossed squares was revealed in WINs compared to that in the controls ($p < 0.05$) and WIN-LDs ($p < 0.01$) (Figure 1). Significant differences were revealed between WIN-LDs and WIN-SDs ($p < 0.05$). An increased total duration of self-grooming was revealed in WIN-SDs and WIN-LDs compared to that in the controls (p < 0.05, p < 0.001, respectively) and WINs (p < 0.05, p < 0.001, respectively).

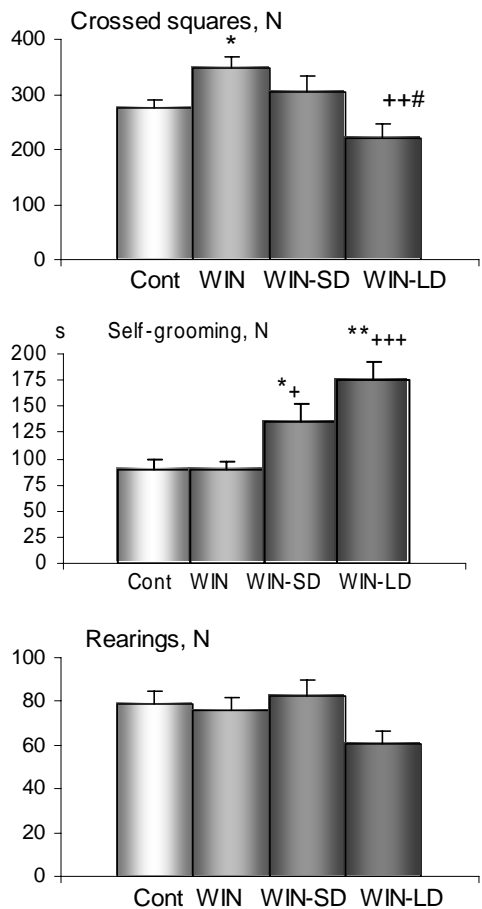

Figure 1. Number of crossed squares, rearings and self-grooming durations in the open-field test in the controls (Cont), winners not exposed to aggression deprivation (WIN), winners exposed to shorter-time aggression deprivation (WIN-SD) and winners exposed to longer-time aggression deprivation (WIN-LD). Scores were obtained before injections. *, p < 0,05, **, p < 0.001 vs Cont; +, p < 0.05; ++, p < 0.01 +++, p < 0.001 vs WIN; #, p < 0.05 vs WIN-SD.

**Behavioral effects of MOR on the winners in the open-field test**

The data are presented in Figure 2. ANOVA revealed the effects of interactions between time intervals and treatment [$F(5,290) = 4.93$, $p < 0.001$] and between time intervals, treatment and experimental groups [$F(15,290) = 1.75$, $p < 0.05$] on the number of crossed squares. A significantly increased number of crossed squares within the 15 – 20-, 20 – 25- and 25 – 30-min time intervals was revealed in the MOR-treated controls compared to saline-treated ones ($p < 0.05$, $p < 0.001$, $p < 0.001$, respectively). There were no significant differences between saline- and MOR-treated WINs and saline- and MOR-treated WIN-SDs in the number of crossed squares ($p > 0.05$ at any point of measurement). The psychomotor activity of 40 % of WIN-LDs (4/10) was not stimulated by MOR: the number of crossed squares was similar to that in the saline-treated WIN-LDs. 60 % of MOR-treated WIN-LDs had a significantly increased number of crossed squares within the 5 – 10-, 15 – 20-, 20 – 25- and 25 – 30-min time intervals compared to the saline-treated WIN-LDs ($p < 0.05$, $p < 0.01$, $p < 0.01$, $p < 0.01$, respectively). No MOR-treated animal in any experimental group displayed self-grooming or rearings.

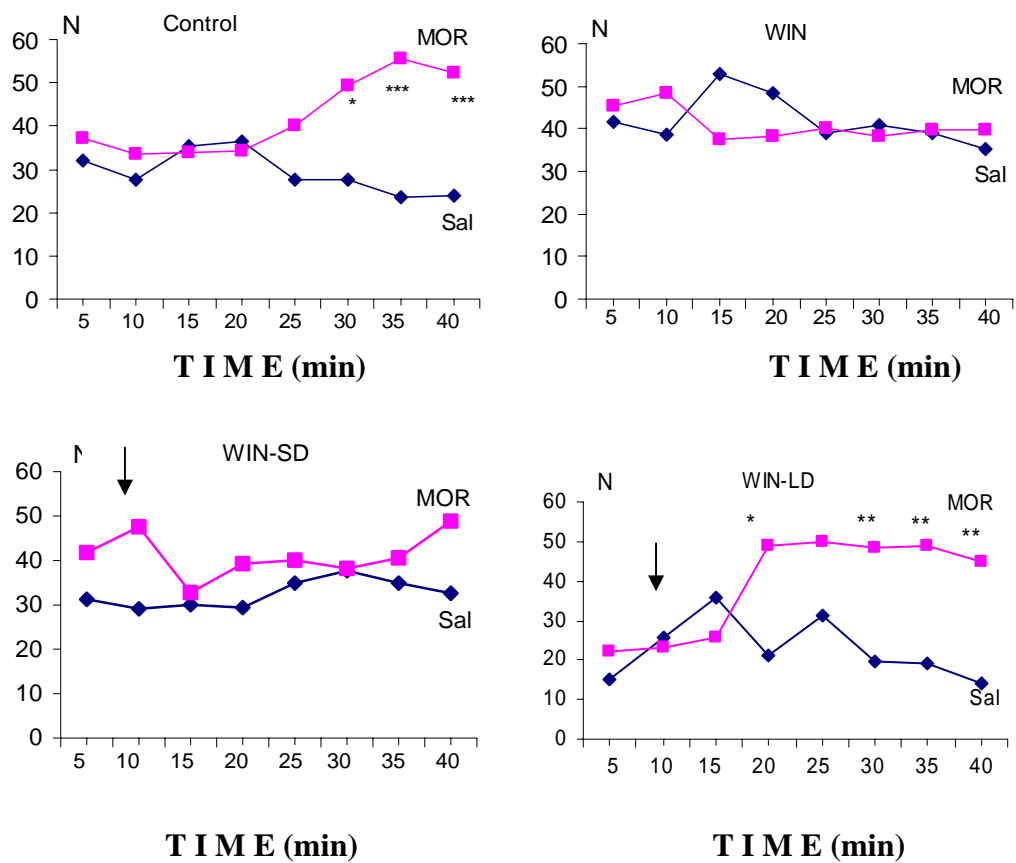

Figure 2. Number of crossed squares in the open-field test after morphine (MOR) and saline (Sal) injections in the controls (Cont), winners not exposed to aggression deprivation (WIN), winners exposed to shorter-time aggression deprivation (WIN-SD) and winners exposed to longer-time aggression deprivation (WIN-LD). *, p < 0.05, **, p < 0.01, ***, p < 0.001 vs saline. Arrow – time of injection.

**Behavioral observations: an elevated tail**

Winners in all groups: 60 % of WINs (12/20), 71 % of WIN-SDs (15/21) and 45 % of WIN-LDs (9/20), demonstrated rigid elevated tails in the open-field test before injection. Tail position was approximately $45°–90°$ relative to the body. The control animals did not raise their tails at that time. MOR injection caused tail elevation for 1 – 2 h in all animal groups. For most of that time, tail position after MOR injection was $45°–90°$ relative to the body in the controls and $90°–180°$, in the winners. A significant difference in the time elapsed between injection and explicit tail elevation was observed between the winners and the controls (Figure 3). One-way ANOVA revealed the effect of experimental group membership on this response time [$F(3.29) = 11.6$, $p < 0.001$]. Shorter response times were observed in WINs, WIN-SDs and WIN-LDs than in the controls (for all $p < 0.001$).

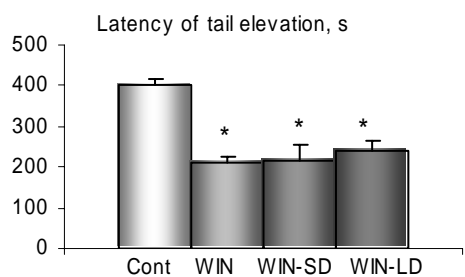

Figure 3. Time intervals between MOR injection and tail elevation in the controls, winners not exposed to aggression deprivation (WIN), winners exposed to shorter-time aggression deprivation (WIN-SD) and winners exposed to longer-time aggression deprivation (WIN-LD). *, p < 0.001 vs Cont.

**Agonistic interaction test**

The Student t-test for dependent data did not reveal any differences in attack variables between the winners before and after 5-day deprivation of aggression (Table).

Table. Level of attacks in the winners after 5 and 14 days of aggression deprivation

| Parameters | Before | Post | P |
|---|---|---|---|
| 5 days of deprivation, n=15 | | | |
| Latency, s | 80,5 ± 33,1 | 84,0 ± 26,7 | > 0,05 |
| Number of attacks | 16,3 ± 3,9 | 15,5 ± 2,9 | > 0,05 |
| Total time of attacks, s | 48,1 ± 9,0 | 58,5 ± 10,6 | > 0,05 |
| Average time of attacks, s | 3,3 ± 0,9 | 5,7 ± 2,2 | > 0,05 |
| 14 days of deprivation, n=11 | | | |
| Latency, s | 122,5 ± 34,7 | 39,2 ± 13,5 | **= 0,02** |
| Number of attacks | 13,6 ± 3,0 | 25,1 ± 6,3 | **= 0,04** |
| Total time of attacks, s | 47,6 ± 15,0 | 80,3 ± 18,3 | > 0,05 |
| Average time of attacks, s | 7,8 ± 5,3 | 3,4 ± 0,3 | > 0,05 |

However, latency to the first attack was significantly shorter and the number of attacks, higher in the winners after than before 14-day deprivation of aggression ($p < 0.05$ and $p < 0.05$, respectively). Behavioral data analysis suggests that, in 60 % of WIN-SDs (9/15) and 80 % of WIN-LDs (9/11), the total attack duration was longer after than before deprivation. Behavioral data on the winners with total attack durations increased after deprivation were re-analysed. Post-deprivation effects were revealed in WIN-SDs and WIN-LDs ($p < 0.01$) (Figure 4).

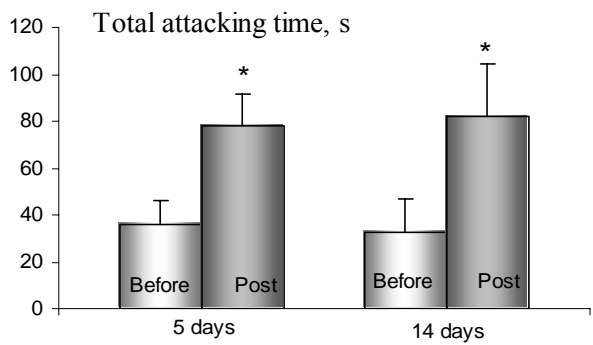

Figure 4. Total attack durations in the winners with increased aggression levels before and after deprivation for 5 and 14 days. *, $p < 0.01$.

**Discussion**

It is known that opioidergic and dopaminergic brain systems are involved in the reward processes and various kinds of social interactions mediating positive emotional responses [34-37], particularly in the mechanism of aggressive behavior [38, 39]. Our experiments have demonstrated that the brain dopaminergic systems have a role in the consequences of repeated aggression experience. Winning a fight causes total activation of the brain dopaminergic systems through an increased dopamine (DA) turnover, which leads to DOPAC (3,4-dehydroxyphenyleacetic acid) formation [40]: elevated DOPAC levels or increased DOPAC/DA ratios in the olfactory bulbs, amygdala, hippocampus, nucleus accumbens, striatum and midbrain were observed in the winners who had 10-day positive fighting experience compared to those in the controls. It has also been shown that dopamine receptors are involved in the development of aggressive behavior [41, 42]. Long-term activation of dopaminergic systems in the winners promotes increase in mRNA levels of tyrosine hydroxylase and DA transporter in the ventral tegmental area (VTA) [43]. Since mesolimbic dopaminergic projections from the VTA play an important role in the mediation of the rewarding processes [for review, see 34], it is possible that the observed changes of gene expression in the VTA of the winners are due to positive emotions from social victories. It has been reported elsewhere that the dopaminergic systems in aggressive rats can be activated: elevated dopamine levels were demonstrated for the prefrontal cortex during and after fights [44].

Tail elevation, which was similar to a morphine-induced Straub tail response, displayed by winners in no matter what experimental conditions, suggests that the opioidergic systems are activated. This hypothesis is supported by the fact that 20-day winners developed tolerance (in some tests, sensitization) to the opioid receptor antagonist naltrexone [21, 22], the kappa-opioid receptor agonist U-50,488H [24] and the mu-opioid receptor agonist DAGO [23].

Thus, there is an apparent similarity between changes in dopaminergic and opioidergic activities in the brain due to repeatedly displaying aggression [25] and the mechanisms of drug addiction [34-37]. Direct confirmation was obtained from exposure of 20-day winners to the sensitization procedure. Behavioral sensitization is characterized by a progressive increase of a psychomotor behavior as a result of repeated administration of a drug [45].

WINs had higher behavioral activity (an increased number of crossed squares) in the open field than the controls. However, WIN-LDs had a decreased number of crossed squares (possibly due to enhanced self-grooming) compared to the controls and WINs. Here enhanced self-grooming may be considered as enhanced replacement activity triggered by deprivation of aggression.

60 % of WIN-SDs and 80 % of WIN-LDs displayed higher levels of aggression after than before deprivation. This enhanced aggression can be regarded as a deprivation effect similar to that after alcohol abuse [46, 47] or drug abuse [for review, see 45]. It was therefore hypothesized that the WIN-SDs and, especially, WIN-LDs had developed addictive states as a result of the altered brain neurochemistry similar to that in drug addicts. This hypothesis was confirmed by the results of the sensitization procedure.

In the controls, the mu-opioid receptor agonist MOR stimulated behavioral activity in the open field: the number of crossed squares was increased, with a maximum score between min 15 and min 30 after injection. In contrast to the saline-treated animals in the respective groups, neither WINs nor WIN-SDs responded to MOR by increased psychomotor activity: it is possible that their mu-opioid receptors became tolerant after prolonged activation of the opioidergic systems due to multiple social victories. In 60 % of WIN-LDs, deprivation of aggression enhanced the effect of psychomotor stimulation by the mu-opioid agonist, which became obvious as early as between min 5 and min 10 after injection and remained increased during the entire observation. The comparison of the psychomotor response to MOR by WIN-LDs and the controls in the open field provides support to the hypothesis that the WIN-LDs became sensitized to the activating effects of MOR.

Time intervals between MOR injection and tail elevation were significantly shorter in the winners of all groups than in the controls. It is therefore possible that multiple aggression experiences cause the development of sensitization of the mu-opioid receptors involved in the regulation of the MOR-induced tail elevation response. As was mentioned above, such sensitization or tolerance of the opioid receptors involved in the control of aggression and associated behaviors was earlier established for another mu-opioid receptor agonist, DAGO, in other tests [23]. In the plus-maze and partition tests, DAGO produced anxiogenic effects on intact mice and was ineffective with 20-day winners, whence it was concluded that repeated aggression experience reduced the sensitivty of the mu-opioid receptors. DAGO increased aggressive grooming in 20-, but not 3-day winners.

**Concluding remarks**. According to Robinson and Berridge [45], the attributes of the development of addiction are long-lasting changes in those brain systems that are normally involved in the process of incentive motivation and reward. These brain reward systems are hypersensitive to drugs or drug-associated stimuli and mediate a subcomponent of reward from wanting to consume addictive drugs. Additionally, the possible causes of addiction are pleasure from consuming a drug and withdrawal symptoms as well as aberrant learning and, especially, the development of strong stimulus-response habits.

The hypothesis that addiction to aggression can be developed due to repeated experience of displaying aggression [25] has now obtained experimental confirmation. Following are possible mechanisms of how the winners develop this addiction.

Aggressive behavior is rewarding both to animals and humans. Psychological studies suggest that engagement in aggressive behavior can be a source of pleasure [48]. Repeated experience of displaying aggression leading to victories causes male mice to develop an aggressive type of behavior (learned aggression) and the intent to be aggressive [18]. In all experimental situations, the experienced winners demonstrate hostile behavior and very seldom do they run away or show defense behavior, even if confronted with a much heavier and stronger mouse. Aggression displayed by some winners is extremely violent and may not be corrected by situational factors; for that reason behavior becomes nonadaptive, pathological [49].

Repeated manifestation of aggression and a graduate acquisition of winning experience induce multiple long-term changes in the mediator systems in the brain [18, 25, 49], which contribute to sensitization to the psychomotor stimulant and motivational effects of aggression. In the winners, the balance between the activity of the mediator systems is disturbed as the excitation processes begin to dominate over inhibitory processes. This disbalance is due to a reduced activity of the serotonergic system in the brain and an enhanced activity of the dopaminergic systems, for example, in VTA and nucleus accumbens [25, 49]. In the winners, chronic activation of the opioidergic systems in the brain causes the development of tolerance to opioid drugs. This reflects a possible development of tolerance to endogenous opioids (for example, brain endogenous morphine existing in the brain [26-28]) or opioid deficiency in the brain.

The winners kept away from fighting for at least 14 days developed sensitization to the activating effects of the mu-opioid receptor agonist MOR. These mice had considerably higher levels of aggression after than before deprivation. Similarly to how it occurs in drug addicts, neuroadaptation for addiction to aggression makes these brain reward systems hypersensitive to aggression-associated stimuli. It might well be that repeated aggression and social victories make the organismal mechanisms that normally regulate aggressive behavior misregulate it, hence a pathology. This pathology, essentially based on the neurochemical disturbances in the brain and aberrant learning, increases aggression in the winners. Under certain circumstances, the effects of endogenous opioids may be abrogated and emotional and physical discomfort may ensue, which eventually leads to internal drive (lust) for aggression, which can result in an outbreak of aggression or seeking out an occasion for behaving aggressively.

There are even more similarities between the mechanisms of addiction to aggression in the victorious mice and in drug addicts than it may seem at first glance: drugs of abuse impact substantially the same neural systems that affect aggressive behavior [50]. It is also well known that drug withdrawal is accompanied by enhanced aggression in animals [30, 31, 51-54] and humans [55-59]. It has been proposed that there are common neurobiological mechanisms underlying reward processes and pleasure phenomena supported by positive emotions of different origins [60]: for example, drug consumption, social success or winning a fight. In the latter two cases, the possible signaling molecules could be endogenous opioids, such as morphine and codeine, which are reportedly present in the brain of various animal species [26-28, 61].

**Acknowledgements**

Author is in debt to Vladimir Filonenko for a thorough revision of this manuscript.